\documentclass[a4paper,11pt]{article}
\usepackage{pos}

\newcommand{\trento}{{\rm T\raisebox{-0.5ex}{R}ENTo}}

\title{Effects of multi-scale jet-medium interactions on jet substructures}

\author*[a]{Yasuki Tachibana}
\onbehalf{for the JETSCAPE Collaboration}

\affiliation[a]{Akita International University, Yuwa, Akita-city 010-1292, Japan}

\emailAdd{ytachibana@aiu.ac.jp}

\abstract{
We utilize event-by-event Monte Carlo simulations within the JETSCAPE framework to examine scale-dependent jet-medium interactions in heavy-ion collisions. The reduction in jet-medium interaction during the early high-virtuality stage, where the medium is resolved at a short distance scale, is emphasized as a key element in explaining multiple jet observables, particularly substructures, simultaneously. By employing the MATTER+LBT setup, which incorporates this explicit reduction of medium effects at high virtuality, we investigate jet substructure observables, such as Soft Drop groomed observables. When contrasted with existing data, our findings spotlight the significant influence of the reduction at the early high-virtuality stages. Furthermore, we study the substructure of gamma-tagged jets, providing predictive insights for future experimental analyses. This broadens our understanding of the various contributing factors involved in modifying jet substructures.
}

\FullConference{
 26-31 March 2023 \\
 Aschaffenburg, Germany\\}

\begin{document}
\maketitle

\section{Introduction}
In their shower evolution, jets exhibit variation in the virtualities and energies of the partonic constituents. Consequently, in high-energy heavy-ion collisions, jets can serve as dynamic probes for investigating jet-medium interactions across different scales. JETSCAPE is a publicly accessible software framework for Monte Carlo event generators that facilitates simulations capturing the physics at various scales involved in in-medium jet evolution. The JETSCAPE framework incorporates multiple models, each effective within a distinct scale range, transitioning between them at appropriate scales while ensuring their intercommunication~\cite{Putschke:2019yrg,JETSCAPE:2019udz,JETSCAPE:2020shq,JETSCAPE:2020mzn,JETSCAPE:2021ehl:qhatfixed,JETSCAPE:2022cob,JETSCAPE:2022jer,JETSCAPE:2022hcb,JETSCAPE:2023hqn}. A notable feature of JETSCAPE is its support for jet quenching strength with explicit virtuality dependence, reflecting the resolution scale evolution of jets. In these proceedings, we demonstrate that this added extension is vital for a comprehensive, concurrent description of the medium modification of jet substructures as well as inclusive jet and single particle spectra. We further delve into a more detailed discussion of the jet substructure modification by focusing on gamma-tagged jets.

\section{Multi-stage jet evolution within the JETSCAPE framework}
We perform simulations using the MATTER+LBT model, a multi-scale in-medium jet evolution model, which can be constructed using the JETSCAPE framework. In this setup, the MATTER module handles the vacuum-like shower evolution of partons with large virtuality, incorporating minor corrections due to the medium effect~\cite{Majumder:2013re,Cao:2017qpx}. Then, the LBT module handles radiation from partons with low virtuality, primarily driven by scatterings with medium constituents~\cite{Wang:2013cia,He:2015pra,Cao:2016gvr}. The switching between the modules is done bidirectionally on a per-parton basis depending on the parton virtuality. Additionally, when the virtuality is high, the transverse size of partons decreases, which causes the QGP to be resolved at a shorter distance scale, appearing more dilute and resulting in fewer interactions~\cite{Kumar:2019uvu}. These virtuality-dependent reductions of jet-medium interaction are incorporated in the MATTER phase in the form of the effective jet quenching strength $\hat{q}_\mathrm{HTL} \cdot f(Q^2)$ with the virtuality-dependent modulation factor parametrized as~\cite{JETSCAPE:2022jer}
\begin{align}
  f(Q^2) & = 
  \begin{cases}
  \frac{1+10\ln^{2}(Q^2_\mathrm{sw}) + 100\ln^{4}(Q^2_\mathrm{sw})}{1+10\ln^{2}(Q^2) + 100\ln^{4}(Q^2)} & 
  \text{for } Q^2 > Q_{\rm sw}^2 \\ 
  1 &\ \text{for } Q^2 \le Q_{\rm sw}^2 
  \end{cases}, 
  \label{eq:qhatSuppressionFactor}
  \end{align}
where $\hat{q}_\mathrm{HTL}$ is the transport coefficient for a low virtuality parton from the conventional hard-thermal-loop (HTL) calculation~\cite{He:2015pra}, $Q^2$ is the virtuality of the jet parton, and $Q^2_\mathrm{sw}$ is the switching virtuality between the MATTER and LBT modules. 
In both modules, the scatterings of jet partons with medium particles are simulated, and the recoiled partons are assumed to be on-shell as they subsequently evolve through the LBT module.

In this study, jet events in PbPb collisions at $\sqrt{s_{\mathrm{NN}}}=5.02~{\rm TeV}$ are generated using the JETSCAPE package. The space-time medium profile for the energy loss calculations is obtained from the $(2+1)$-D evolution of freestreaming~\cite{Liu:2015nwa} and subsequent viscous hydrodynamics by VISHNU~\cite{Shen:2014vra}, with the geometrical initial condition provided by \texttt{\trento}\ \cite{Moreland:2014oya}. The hard partons are generated by PYTHIA8~\cite{Sjostrand:2019zhc}, and their initial spatial conditions are sampled using the same \texttt{\trento}\ initial conditions as those of the medium. These partons undergo in-medium shower evolution by MATTER+LBT with a switching virtuality $Q^2_\mathrm{sw}=4~\mathrm{GeV}^2$, and are subsequently hadronized by the Lund string model implemented in PYTHIA 8.

\section{Results}

\begin{figure*}
\begin{center}
  \includegraphics[width=0.48\textwidth]{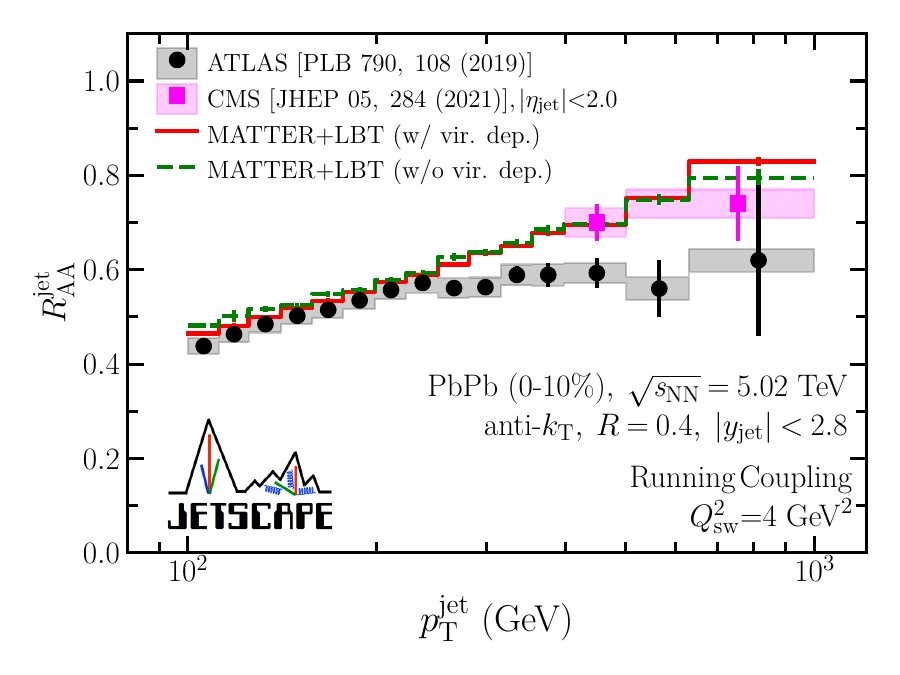}
  \includegraphics[width=0.48\textwidth]{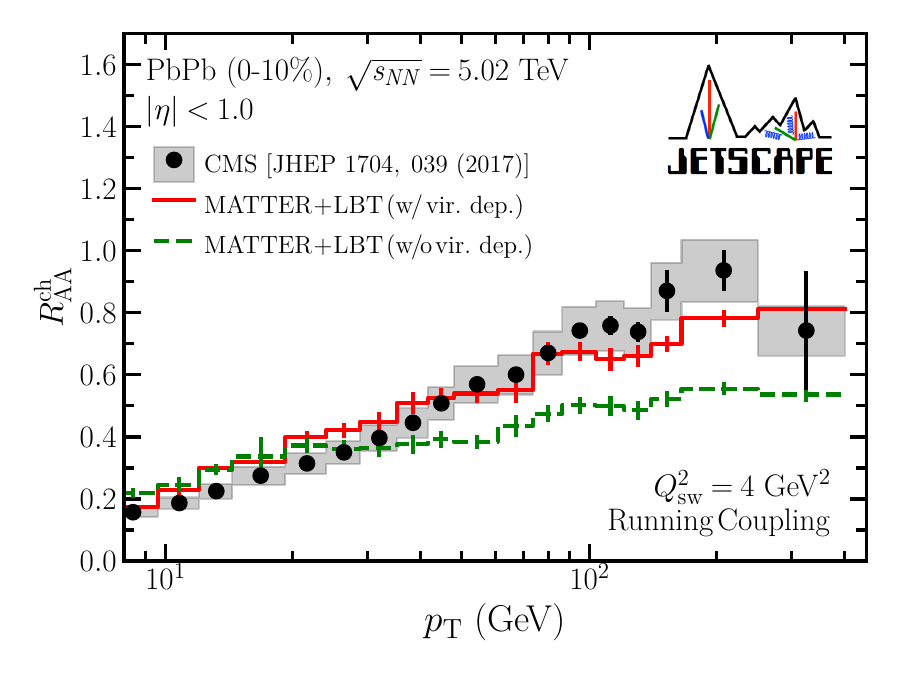}  
\vspace{-8.5pt}
\caption{Nuclear modification factors ($R_{\mathrm{AA}}$) for reconstructed jets (left) and single charged particles (right) in 0-10\% Pb+Pb collisions at $5.02$~TeV. The MATTER+LBT results with (solid red) and without (dashed green) the virtuality dependence of the modulation factor~(\ref{eq:qhatSuppressionFactor}) are compared. Experimental data are taken from ATLAS~\cite{ATLAS:2018gwx} and CMS~\cite{CMS:2021vui,CMS:2016xef}.}
\label{raa}
\end{center}
\end{figure*}
Figure~\ref{raa} shows the nuclear-modification factor for the reconstructed jets and single charged particles. Here, two different setups are compared: with (solid red) and without (dashed green) the explicit virtuality dependence via the modulation factor~(\ref{eq:qhatSuppressionFactor}). The parameters are tuned separately for each setup, with the former using an $\alpha_{\mathrm{s}} = 0.3$, and the latter using $f(Q^2)=1$ across all ranges of $Q^2$, with an $\alpha_{\mathrm{s}} = 0.25$. While no apparent difference between these two setups can be observed in the reconstructed jet suppression, clear sensitivity is evident in the single particle suppression, with the rising trend in the experimental data being reproduced only in the setup incorporating the virtuality dependence.

\begin{figure*}
\begin{center}
\includegraphics[width=0.48\textwidth]{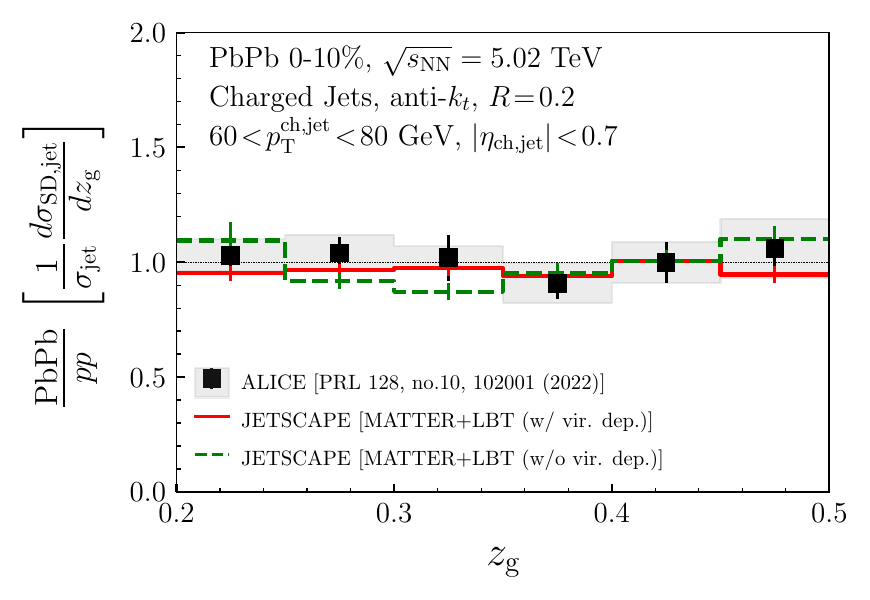}
\includegraphics[width=0.48\textwidth]{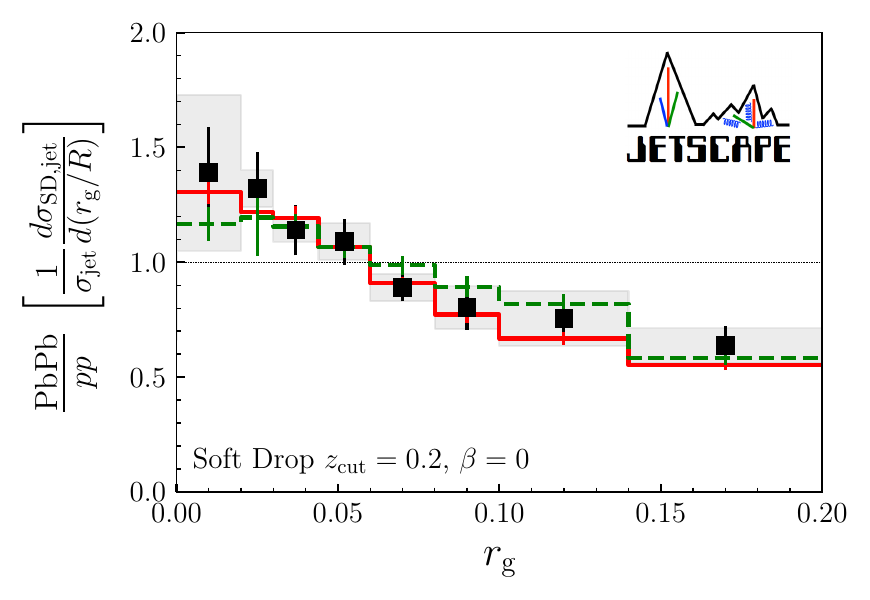}  
\vspace{-8.5pt}
\caption{PbPb/pp ratios of inclusive charged jet $z_g$ (left) and $r_g$ (right) distributions for the 0-10 \% centrality at $5.02$~TeV. Experimental data are taken from ALICE~\cite{ALargeIonColliderExperiment:2021mqf}.}
\label{alice_sd}
\end{center}
\end{figure*}
In Fig.~\ref{alice_sd}, we present the results of Soft Drop observables for charged jets. The $z_{g}$ distribution exhibits virtually no observable medium modification, aligning with the ALICE data for both cases with and without the virtuality dependence. Regarding the $r_g$ distribution, considerable differences emerge due to variations in the setup; however, under this configuration, both approaches display largely monotonic behavior, coinciding with the ALICE data within the errors.

\begin{figure*}
\begin{center}
\includegraphics[width=0.95\textwidth]{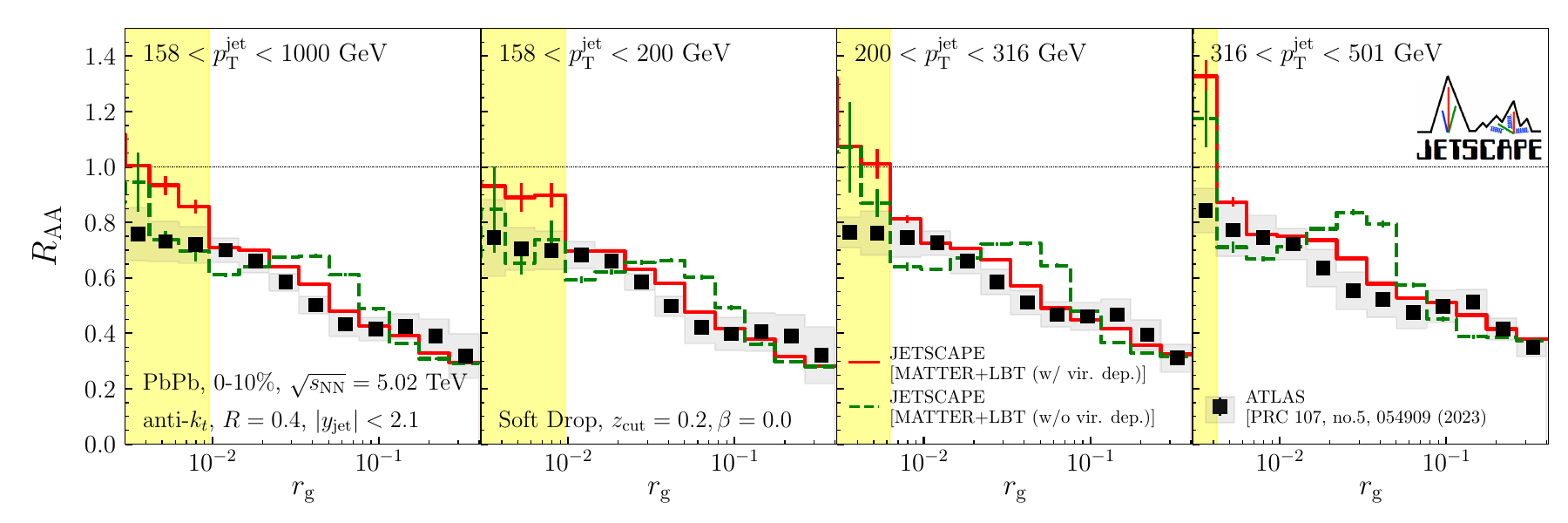}
\vspace{-8.5pt}
\caption{Nuclear modification factors $R_{\mathrm{AA}}$ as a function of $r_{g}$ for inclusive full jets with different $p_{T}^{\mathrm{jet}}$ in 0-10\% Pb+Pb collisions at $5.02$~TeV. Experimental data are taken from ATLAS~\cite{ATLAS:2022vii}. }
\label{atlas_sd}
\end{center}
\end{figure*}
Figure~\ref{atlas_sd} presents the $r_{\mathrm{g}}$-dependent $R_{\mathrm{AA}}$ results for inclusive full jets, compared to ATLAS data~\cite{ATLAS:2022vii}. In this observable configuration, the differences between the two setups become strikingly evident, even at the qualitative behavior level. The experimental data exhibits a monotonic behavior and strongly favors the case with the explicit virtuality-dependent formulation. This implies that scatterings with medium constituents at high virtuality are reduced due to the finer scale probed by the jet parton in the medium.

\begin{figure*}
\begin{center}
\includegraphics[width=0.95\textwidth]{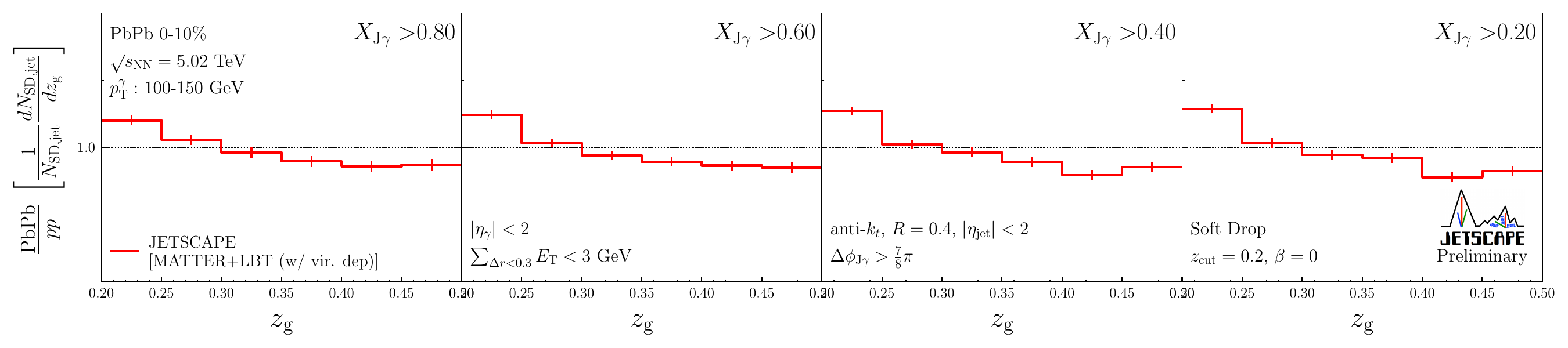}
\includegraphics[width=0.95\textwidth]{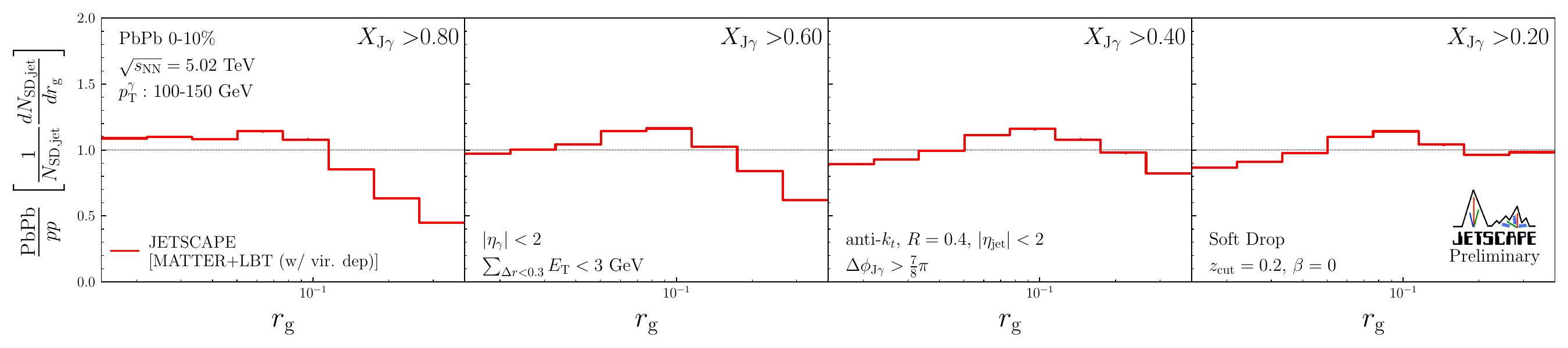}  
\vspace{-8.5pt}
\caption{PbPb/pp ratios of gamma-tagged jet $z_g$ (top) and $r_g$ (bottom) distributions for different ranges of gamma-jet asymmetry $X_{\mathrm{J\gamma}}$ for the 0-10 \% centrality at $5.02$~TeV. }
\label{gamma_sd}
\end{center}
\end{figure*}
In Fig.~\ref{gamma_sd}, we present our predictions for the modification of the $z_{\mathrm{g}}$ and $r_{\mathrm{g}}$ distributions in photon-tagged jets, calculated with the virtuality-dependent modulation factor. Each panel represents a different gamma-jet asymmetry ($X_{\mathrm{J}\gamma}=p^{\mathrm{jet}}_{\mathrm{T}}/p^{\mathrm{\gamma}}_{\mathrm{T}}$) cut, which estimates jet energy loss. The modification trend in the $z_{\mathrm{g}}$ distribution remains consistent regardless of the constraint on $X_{\mathrm{J}\gamma}$, indicating that it is independent of the magnitude of energy loss. On the other hand, the $r_{\mathrm{g}}$ distribution exhibits a clear dependence on energy loss. When the lower limit of $X_{\mathrm{J}\gamma}$ is high, indicating that only jets with minimal energy loss are observed, significant suppression is observed at large $r_{\mathrm{g}}$, which is also observed in the case of inclusive jets (Figs. \ref{alice_sd} and \ref{atlas_sd}). However, this suppression largely disappears when the lower limit of $X_{\mathrm{J}\gamma}$ is reduced, allowing jets with greater energy loss to be included, thereby reducing trigger bias. This suggests that the observed suppression at large $r_{\mathrm{g}}$ in inclusive jets and photon-tagged jets with large $X_{\mathrm{J}\gamma}$ cuts does not indicate a narrowing of the hardest jet splittings due to medium effects. The medium effects only slightly modify the jet $r_{\mathrm{g}}$. Jets with larger inherent $r_{\mathrm{g}}$ experience greater energy loss. As a result, this pronounced suppression of $r_{\mathrm{g}}$ is evident in inclusive jets and photon-tagged jets with large $X_{\mathrm{J}\gamma}$ cuts, but not in photon-tagged jets with smaller $X_{\mathrm{J}\gamma}$ cuts.

\section{Summary}
We presented a study on jet-medium interactions through Monte Carlo simulations using the MATTER+LBT model within the JETSCAPE framework. By incorporating a virtuality-dependent modulation factor, the study demonstrates a comprehensive and concurrent description of reconstructed jet spectra, single-particle spectra, and jet substructure modifications compared to experimental data. The results highlight the significance of the virtuality dependence in accurately capturing the effects of the medium on jets, revealing the reduced interaction of high-virtuality partons with the medium. Moreover, the analysis of photon-tagged jets indicates that the observed suppression at large $r_{\mathrm{g}}$ does not indicate narrowing due to medium effects but rather strongly reflects that jets with larger opening angles are more susceptible to energy loss.

\bibliographystyle{h-physrev3}
\bibliography{skeleton,extrabib}

\end{document}